\documentclass[aps,preprint,amsmath,amssymb]{revtex4}
\usepackage{graphicx}
\usepackage{epstopdf}
\usepackage{color}
\begin{document}
\title{Anomalous $WW\gamma$ couplings with beam polarization at the Compact Linear Collider}

\author{V. Ar{\i}}
\email[]{vari@science.ankara.edu.tr} \affiliation{Department of
Physics, Ankara University, 06100, Ankara, Turkey}

\author{A. A. Billur}
\email[]{abillur@cumhuriyet.edu.tr} \affiliation{Department of
Physics, Cumhuriyet University, 58140, Sivas, Turkey}

\author{S.C. \.{I}nan}
\email[]{sceminan@cumhuriyet.edu.tr}
\affiliation{Department of Physics, Cumhuriyet University,
58140, Sivas, Turkey}

\author{M. K\"{o}ksal}
\email[]{mkoksal@cumhuriyet.edu.tr} \affiliation{Department of
Physics, Cumhuriyet University, 58140, Sivas, Turkey}

\begin{abstract}
We study the anomalous $WW\gamma$ couplings at the Compact Linear Collider through the processes $e^{+}e^{-}\to W^+W^-$,
$e^{-}e^{+} \to e^{-} \gamma^{*} e^{+} \to e^{+} \nu_{e} W^-$ and $e^{-}e^{+}\to e^{-} \gamma^{*} \gamma^{*} e^{+}  \to e^{-} W^+ W^- e^{+} $ $ (\gamma^{*}$ is the Weizsacker-Williams photon). We give the 95\% confidence level limits for unpolarized and polarized electron (positron) beam on the anomalous couplings for various values of the integrated luminosities and center-of-mass energies. We show that the obtained limits on the anomalous couplings through these processes  can highly improve the current experimental limits.  In addition, our limits with beam polarization are approximately two times better than the unpolarized case.
\end{abstract}

\maketitle

\section{Introduction}

The Standard Model (SM) has been so far successful in describing the below the electroweak scale with high precision. Therefore, electroweak interactions are known very well in this model. Self-interactions of the gauge bosons are outcomes of the $SU_L(2)\times U_Y(1)$ gauge symmetry of the SM. Determination of these type of interactions plays an important role to test the non-Abelian gauge symmetries of the electroweak sector. Searching that kind of interactions can generate extra confirmation of the SM with a higher sensitivity or reveal some information for new physics beyond the SM. Any measurement which conflicts with the SM expectations would lead to the existence of new physics.

The traditional approach to investigate new physics effect to $WW\gamma$ interactions introduces in a model independent way by means of the effective Lagrangian method. The theoretical motivations of such a method would be based on the guess that at higher energy regions beyond the SM, there is a main physics which reduces to the SM at lower energy regions.
Such a procedure is quite general and independent of the details of the model. Hence, this method is generally known model independent analysis. The effective Lagrangian for $WW\gamma$ interaction which conserves charge and parity can be given as follows \cite{hag,gae},

\begin{eqnarray}
\frac{iL}{g_{WW\gamma}}=g_1^\gamma(W_{\mu\nu}^\dag W^\mu A^\nu-W^{\mu\nu} W_{\mu}^\dag A_\nu)+\kappa W_\mu^\dag W_\nu A^{\mu\nu}+\frac{\lambda}{M_W^2}W_{\rho\mu}^\dag W_\nu^\mu A^{\nu\rho}.
\label{efl}
\end{eqnarray}

 Here $g_{WW\gamma}=e$ , $V_{\mu\nu}=\partial_\mu V_\nu-\partial_\nu V_\mu$ ($V_\mu=W_\mu, A_\mu$), $g_1^\gamma$, $\kappa$, $\lambda$ are the dimensionless anomalous parameters. They are related to magnetic dipole and electric quadrupole moments of $W$ boson. In the SM, the couplings are obtained  $g_1^\gamma=1$, $\kappa=1$ and $\lambda=0$ at the tree level. The $g_1^\gamma=1$ value is fixed for on-shell photons at tree level by electromagnetic gauge invariance to its SM value. Then, the Feynman rule for the anomalous vertex can be found from Eq.(\ref{efl}),

 \begin{eqnarray}
 \Gamma_{\mu\nu\rho}=&&eg_{\mu\nu}\left(p_1-p_2-\frac{\lambda}{M_W^2}[(p_2.p_3)p_1-(p_1.p_3)p_2]\right)_\rho \nonumber \\
&&+eg_{\mu\rho}\left(\kappa p_3-p_1+\frac{\lambda}{M_W^2}[(p_2.p_3)p_1-(p_1.p_2)p_3]\right)_\nu \nonumber \\
&&+eg_{\nu\rho}\left(-\kappa p_3+p_2-\frac{\lambda}{M_W^2}[(p_1.p_3)p_2-(p_1.p_2)p_3]\right)_\mu \nonumber \\
 &&+\frac{e\lambda}{M_W^2}(p_{2\mu}p_{3\nu}p_{1\rho}-p_{3\mu}p_{1\nu}p_{2\rho})
 \end{eqnarray}
 where all of momentums are incoming the vertex.

However, after the recent discovery of a new particle which is consistent with the SM Higgs boson, then new physics is described in terms of a direct extension of the ordinary SM formalism; i.e. using a linear realization of the symmetry. Considering CP-conserving interactions of dimension six, eleven independent operators can be constructed. Among them the three operators which do not affect the gauge boson propagators at tree-level, but give rise to deviations in the charge and parity conserving $WW\gamma$ gauge couplings. Denoting the corresponding couplings as $\alpha_{W\phi}$, $\alpha_{B\phi}$, and $\alpha_{W}$, the $WW\gamma$ couplings inducing effective Lagrangian can be given by \cite{TGC}

\begin{eqnarray}
L=ig'\frac{\alpha_{B\phi}}{m_{W}^{2}}(D_{\mu}\Phi)^{\dag}B^{\mu\nu} (D_{\nu}\Phi)+ig\frac{\alpha_{W\phi}}{m_{W}^{2}}(D_{\mu}\Phi)^{\dag} \overrightarrow{\tau}.\overrightarrow{W}^{\mu\nu}(D_{\nu}\Phi)+9\frac{\alpha_{W}}{6 m_{W}^{2}}\overrightarrow{W}_{\nu}^{\mu}.(\overrightarrow{W}_{\rho}^{\nu}\times\overrightarrow{W}_{\rho}^{\mu})
\end{eqnarray}

Replacing the Higgs doublet field by its vacuum expectation value in the above equation, nonvanishing anomalous $WW\gamma$ gauge couplings in Eq. 1 can be expressed as

\begin{eqnarray}
\Delta\kappa_{\gamma}= \alpha_{W\phi}+\alpha_{B\phi}, \,\,\,\,\,\,\,\,   \lambda_{\gamma}=\alpha_{W}.
\end{eqnarray}
There are a lot of phenomenological studies for $WW\gamma $ interactions at the linear and hadron colliders \cite{ph1,ph2,ph3,ph4,ph5,ph6,ph7,ph8}. The experimental sensitivity limits on anomalous $WW\gamma$ couplings $\Delta\kappa=\kappa-1$ and $\lambda$  are obtained by the LEP \cite{lep}, Tevatron \cite{tev} and LHC \cite{lhc,cms,atlas}. The obtained results have been shown in Table \ref{tab1} at 95\% confidence level. The best stringent limits on anomalous $WW\gamma$ couplings have been obtained by the LEP. However,  $WW\gamma$ couplings can not be well distinguished from the $WWZ$ couplings in this experiment. The constraints of the LHC bounds on anomalous couplings are significantly greater than that of the Tevatron due to the higher center-of-mass energy and higher $WW$ events. These limits are also comparable to the LEP results.

The linear $e^{+}e^{-}$ collider with high energy and high luminosity can give opportunity to higher precision than the hadron collider. One of the possibilities of new type linear collider is the Compact Linear Collider (CLIC).
CLIC energies span from $0.5$ to $3$ TeV and luminosity up to $ 590 $ $ fb^{-1} $ and we have taken these parameters to conform with \cite{dan,19,dan2}.The linear colliders may have an another option that polarized beam collisions. These type of collisions give new perspectives such as on the hadronic structure and high precision measurements on the electroweak mixing angle \cite{pol1}. Beam polarization could be important role in the next linear colliders as well as RHIC and HERA. It is expected that $80 \%$ polarization of lepton beam can be achievable at the future linear colliders \cite{pol2}. In this work, we take into account one beam can be $\pm 80 \%$ polarization ($+80 \% $ means that eighty of percent are right polarized) and one beam can be -$60 \%$ (this means that sixty of percent are left polarized).

After high energy linear colliders have been constructed, its operating modes of $e \gamma$ and $\gamma \gamma$  \cite{35,36} are expected to be made. Here real photons are obtained by Compton backscattering mechanism. However,
$\gamma ^{*} \gamma^{*}$ and $e \gamma^{*}$ interactions can appear spontaneously with respect to $\gamma \gamma$ and $e \gamma$ interactions \cite{39,40,41,42}. Therefore, $\gamma ^{*} \gamma^{*}$ and $e \gamma^{*}$ collisions are more realistic than the Compton backscattering procedure search for the new physics beyond the SM. These reactions occur with quasi-real photons are emitted from one (or two) of the
lepton beams. These processes can be defined by the Weizsacker-Williams approximation (WWA)\cite{wwa1,wwa2}. In this approximation, the virtuality of the photons are very small. Therefore, scattered angels of the emitting photons from the leptons trajectory along
the actual beam path should be very small. The use of the WWA provides a lot of benefits.
With simple formulas, it let to obtain simple numerical estimations \cite{WWA}. Also, this method
provides a facility in the experimental studies since it allows to give
events number for $\gamma^{*} \gamma^{*}\rightarrow X $ process approximately through the examination of the $e^{-}e^{+}\rightarrow e^{-}Xe^{+}$ scattering \cite{WWA}. Moreover, these processes have a very clean experimental environment, since they have no interference with weak and strong interactions.

There are many phenomenological and experimental analysis about the WWA at the LEP, Tevatron and LHC colliders \cite{sah,399,400,401,402,43,44,45,q1,q2,q3,q4,q5,q6,q7,q8,q9,q10,q11,q12,q13,s6,s7,47b,47c,sx,r1,r2,r3,r4,r5,r6,r7,r8}. Furthermore, many studies on new physics beyond the SM using the WWA at the CLIC in the literature have been phenomenologically investigated. These searches involve: gauge boson self-interactions, excited neutrino, the electromagnetic moments of the tau lepton (neutrino), doubly charged Higgs bosons and so forth \cite{47a,451,47,555,s1,s2,s3,s4,s5}.

In this study, we search for $e^-e^+\to W^+W^-$, $e^-e^{+}\to e^-\gamma^{*}e^{+}\to W^-v_ee^{+}$, $e^{-}e^{+}\to e^{-}\gamma^{*}\gamma^{*}e^{+} \to e^{-} W^-W^+e^{+}$ processes to investigate $WW\gamma$ anomalous couplings. One of the advantages of $\gamma^{*}\gamma^{*} $ and $\gamma^{*} e$ processes is that they can isolate $WW\gamma$ couplings from $WWZ$ couplings in $e^-\gamma^{*}\to W^-v_e$, $\gamma^{*}\gamma^{*} \to W^-W^+$ processes.

There are several terms in tree-level cross section. These are proportional to $\Delta\kappa^2$, $\lambda^2$, $\Delta\kappa$, $\lambda$ and $\Delta\kappa\lambda$  additional to the SM cross section. In the effective Lagrangian, the energy dependence of $\Delta\kappa$ and $\lambda$ terms are different as seen from (Eq.\ref{efl}). Especially, limits on $\lambda$ are stronger than $\Delta\kappa$ (see Table \ref{tab1}). Furthermore, it can be seen from Table \ref{tab1} obtained limits on $\Delta\kappa$ from the hadron colliders are much weaker than lepton colliders. For this reason, lepton colliders open new opportunities to search these anomalous $WW\gamma$ and $WWZ$ couplings.

All numerical calculations related to anomalous $WW\gamma$ interaction vertices can be evaluated via CalcHEP \cite{calchep,48}. This new model can be added five new files by hand into CalcHEP by writing a set of pure text model files which contain all the details of the model including the properties of its particles, parameters and vertex rules. These files are Lagrangian, Variables, Composite, Constraints and Particles that appearing in the model files of CalcHEP. Firstly,  the anomalous $WW\gamma$ vertices defined through the effective Lagrangian given with Eq. (1) are replaced with SM $WW\gamma$ vertices in Lagrangian file according to the interaction vertex rules in Refs. \cite{calchep,48}. Subsequently, $\Delta\kappa$, and $\lambda$ couplings in this effective Lagrangian are defined in Variables file. Other files are not be any change.  Finally, routine of the rules given in Refs. \cite{calchep,48} are performed numerical calculations for the three processes including new physics beyond the SM.

To obtain limits, one-parameter sensitivity analysis we take into account $\chi^2$ test,

\begin{equation}
\chi^{2}=\left(\frac{\sigma_{SM}-\sigma(\Delta\kappa, \lambda)}{\sigma_{SM} \,\, \delta}\right)^{2}
\end{equation}

\noindent where $\sigma(\Delta\kappa, \lambda)$ is the total cross section including $SM$ and new physics, $\delta=1/\sqrt{N}$; $N=\sigma_{SM} BR(W \rightarrow l \nu)L_{int}$. We have used that only one of the anomalous coupling is non zero at any given time, while the other
one anomalous coupling is taken to zero. For the total cross section of the $e^+e^- \to W^+ W^-$ and $\gamma^{*} \gamma^{*} \to W^+ W^-$ processes, we assume that one of the bosons decays is hadronic and the other is leptonic. For these processes, we take into account $BR=0.145$. For the process $ e^{-} \gamma^{*} \to W^{-} \nu_{e}$, we assume hadronic decay channel $BR=0.67$.

\section{Numerical Analysis}
\subsection{Anomalous couplings via  the process $e^{+}e^{-} \to W^{+} W^{-}$ }

\begin{figure}
\includegraphics{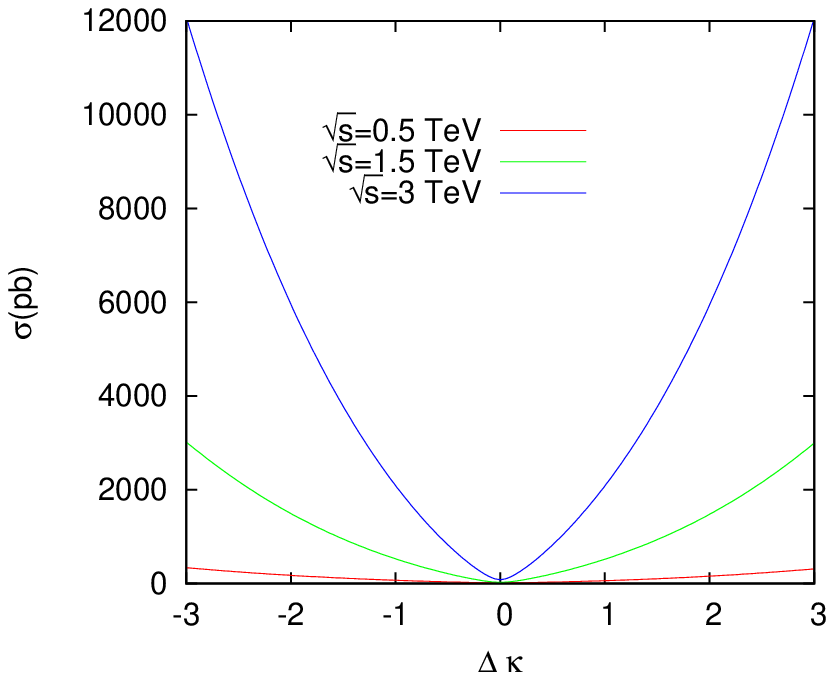}
\caption{The total cross section for process  $e^+e^- \to W^+ W^- $ as a function of $\Delta\kappa$ at unpolarized case and various values of center-of-mass energy.
\label{figcsk}}
\end{figure}

\begin{figure}
\includegraphics{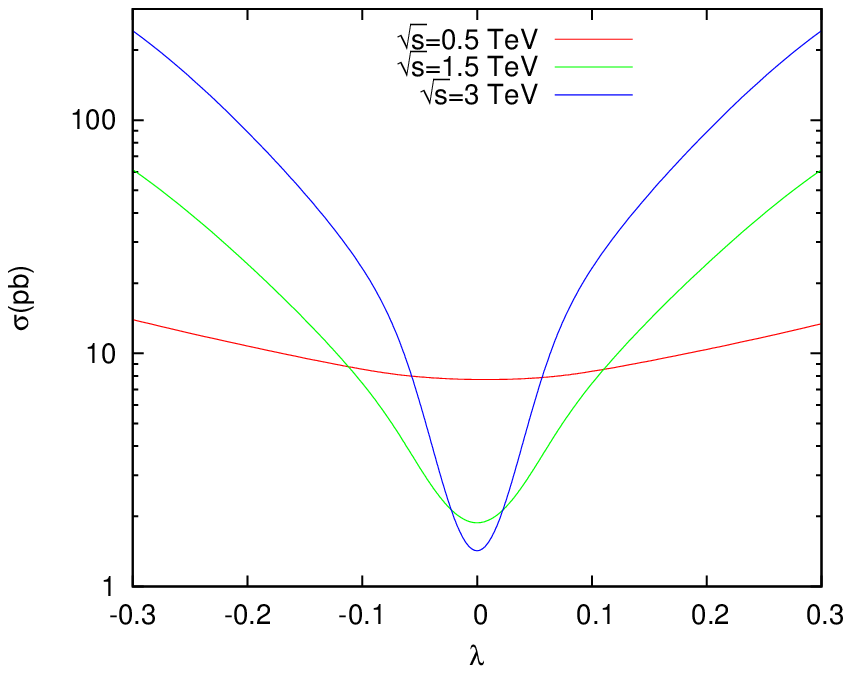}
\caption{The total cross section for process  $e^+e^- \to W^+ W^-$ as a function of $\lambda$ at unpolarized case and various values of center-of-mass energy.
\label{figcsl}}
\end{figure}

We examine $e^+e^- \to W^+ W^- \to q\bar{q'}lv$ ($q;q'=u,d,s;l=e,\mu$) process at the CLIC energies search for anomalous $WW\gamma$ couplings. We have made analysis for both  the unpolarized and polarized electron and positron. In Figs.\ref{figcsk} and \ref{figcsl}, we obtain the cross sections as functions of the anomalous couplings $\Delta\kappa$ and $\lambda$ for the unpolarized cases with using three center-of-mass energies $\sqrt{s}=0.5, 1.5$ and $3$ TeV.  The cross sections seem highly
depend on the center-of-mass energies and anomalous couplings. The cross-sections have a quadratic dependence on $\Delta\kappa$ and $\lambda$ since new physics contribution including $ \Delta\kappa^2$ and $\lambda^2$ terms a lot more at high energy region. Therefore, although anomalous couplings are very small, the contribution proportional to the quadratic terms are not negligible. Moreover the cross sections are symmetric under the sign of the anomalous couplings. Therefore, the interference terms are not dominant.

\begin{figure}
\includegraphics{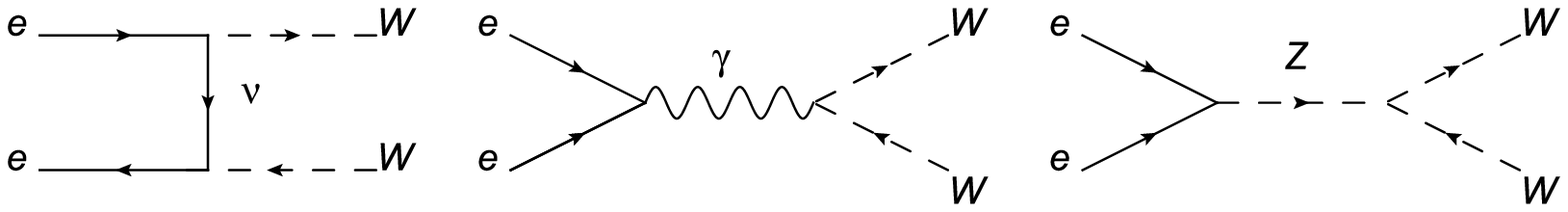}
\caption{Tree-level Feynman diagrams for the subprocess $e^+e^- \to W^+ W^- $ .
\label{figeediagram}}
\end{figure}

\begin{figure}
\includegraphics{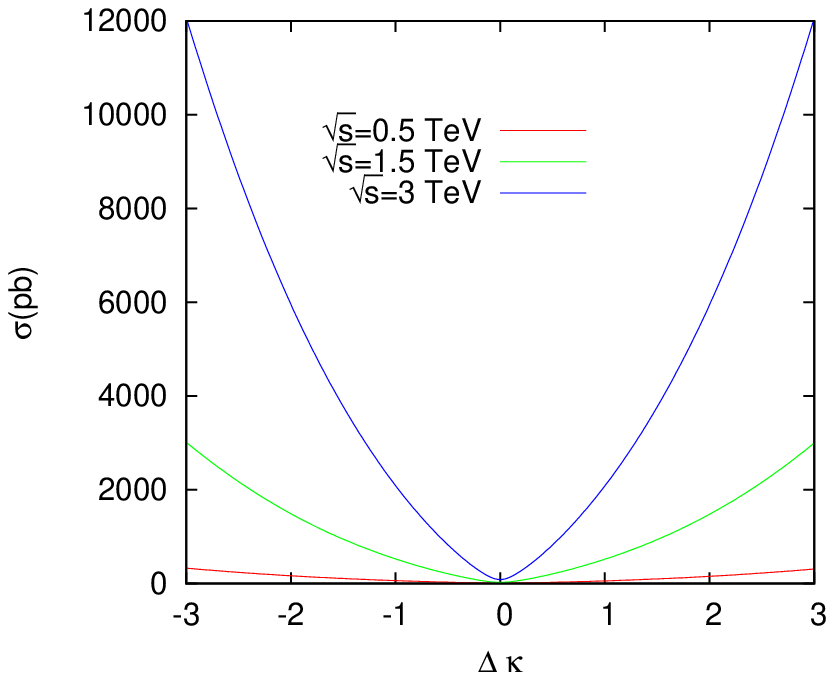}
\caption{The total cross section for process $e^+e^- \to W^+ W^- $ as a function of $\Delta\kappa$ at the $\sqrt{s}=0.5, 1$ and $3$ TeV with $P(e^-)=80\%$ and $P(e^+)=0\%$ polarizations.
\label{figcskpol1}}
\end{figure}

\begin{figure}
\includegraphics{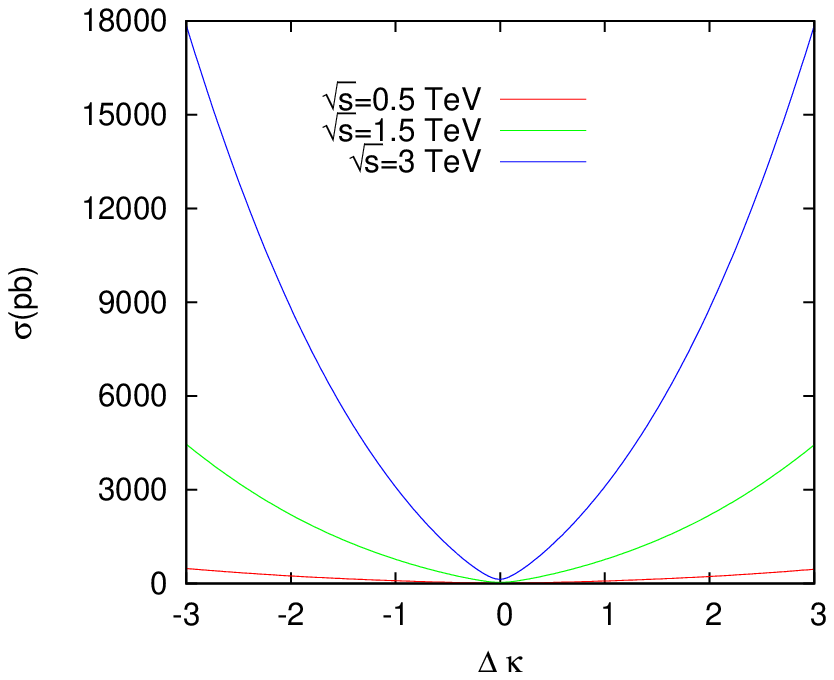}
\caption{The total cross section for process $e^{-}e^{+}\rightarrow W^{-}W^{+}$ as a function of $\Delta\kappa$ at the $\sqrt{s}=0.5, 1$ and $3$ TeV with $P(e^-)=80\%$ and $P(e^+)=$-$60\%$ polarizations.
\label{figcskpol2}}
\end{figure}

\begin{figure}
\includegraphics{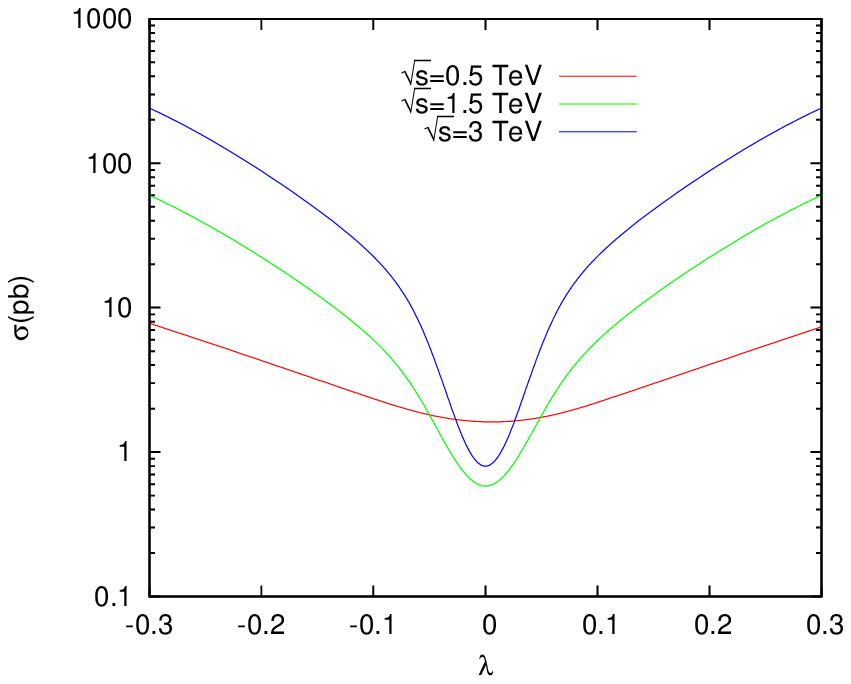}
\caption{The total cross section for process $e^+e^- \to W^+ W^-$ as a function of $\lambda$ at the $\sqrt{s}=0.5, 1$ and $3$ TeV with $P(e^-)=80\%$ and $P(e^+)=0\%$ polarizations.
\label{figcslpol1}}
\end{figure}

\begin{figure}
\includegraphics{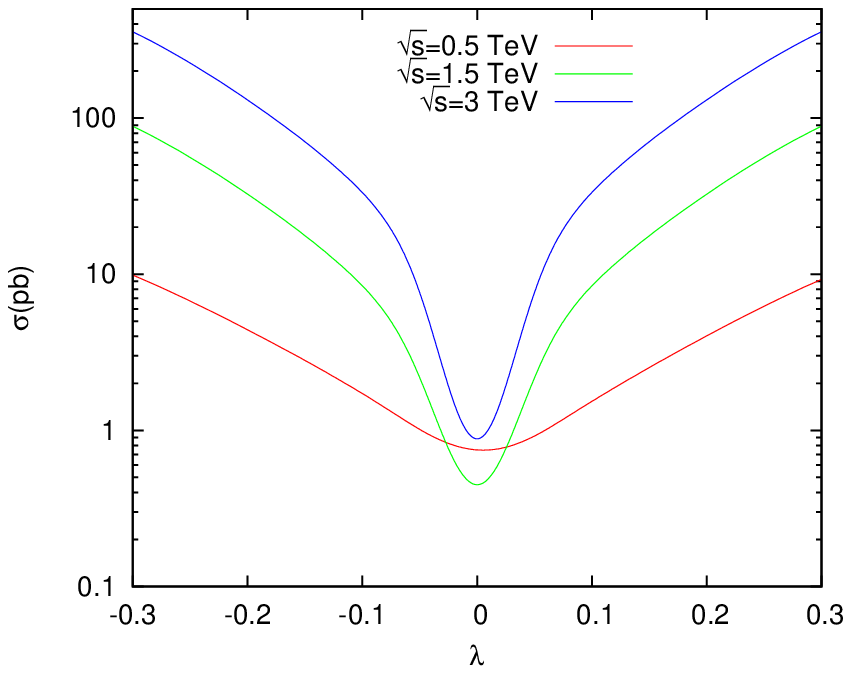}
\caption{The total cross section for process $e^{-}e^{+}\rightarrow W^{-}W^{+}$ as a function of $\lambda$ at the $\sqrt{s}=0.5, 1$ and $3$ TeV with $P(e^-)=80\%$ and $P(e^+)=$-$60\%$ polarizations.
\label{figcslpol2}}
\end{figure}

For impose an idea about the effect of the unpolarized and polarized electron (positron) cases, we represent the  total cross sections in Figs.\ref{figcskpol1} and \ref{figcskpol2} as functions of $ \Delta\kappa $ for $\sqrt{s}=0.5, 1.5$ and $3$ TeV. We take the electron beam polarization $P(e^-)=80\%$, positron beam polarization $P(e^+)=0 \%$ in Fig.\ref{figcskpol1} and $P(e^-)=$-$80\%$, $P(e^+)=$-$60\%$ in Fig.\ref{figcskpol2}. In Fig.\ref{figeediagram}, there are three diagrams at the tree level. Second of these includes $WW\gamma $ vertex and it gives the most contribution to the total cross section. For the $P(e^-)=80 \%$ and $P(e^+)=0 \%$ polarization in Fig.\ref{figcskpol1}, total cross section does not change significantly. The cause of this condition is that the second diagram contribution does not notable change, first and third diagrams contribution is less than the second one to the total cross section in this polarization state. For $P(e^-)=80\%$ and $P(e^+)=$-$60\%$ case, contribution of the diagram which includes new physics interactions is changed too large. It can be seen from Fig. \ref{figcskpol2} that polarization of the $e^+$ ($e^-$) beams strongly changes to the total cross section as a variation of $\Delta\kappa$. Similar behavior for the total cross section can be seen in Figs.\ref{figcsl}, \ref{figcslpol1} and \ref{figcslpol2} for $\lambda$ anomalous coupling.

\begin{figure}
\includegraphics{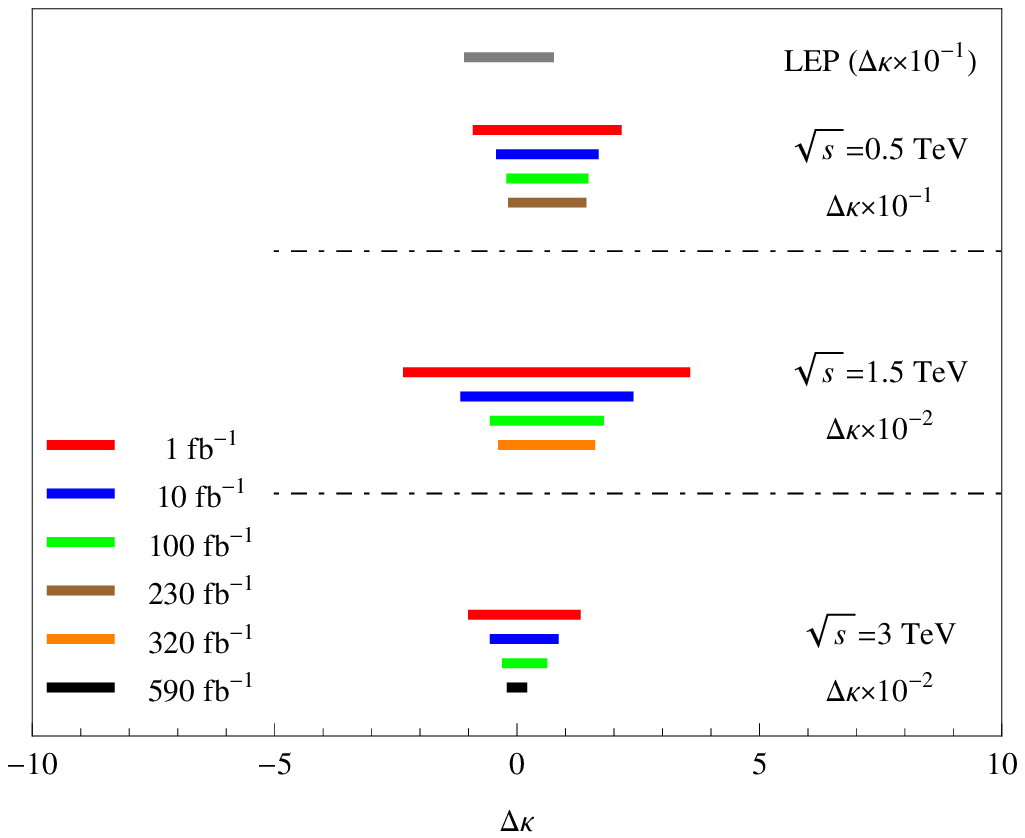}
\caption{95\% C.L. sensitivity limits of the  $\Delta\kappa$ coupling
for various values of integrated CLIC luminosities and center-of-mass energies. $e^{-}e^{+}\rightarrow W^{-}W^{+} \to q\bar{q'}lv$ ($q;q'=u,d,s;l=e,\mu$) processes with unpolarized beams have been considered.}
\label{figlimk}
\end{figure}

\begin{figure}
\includegraphics{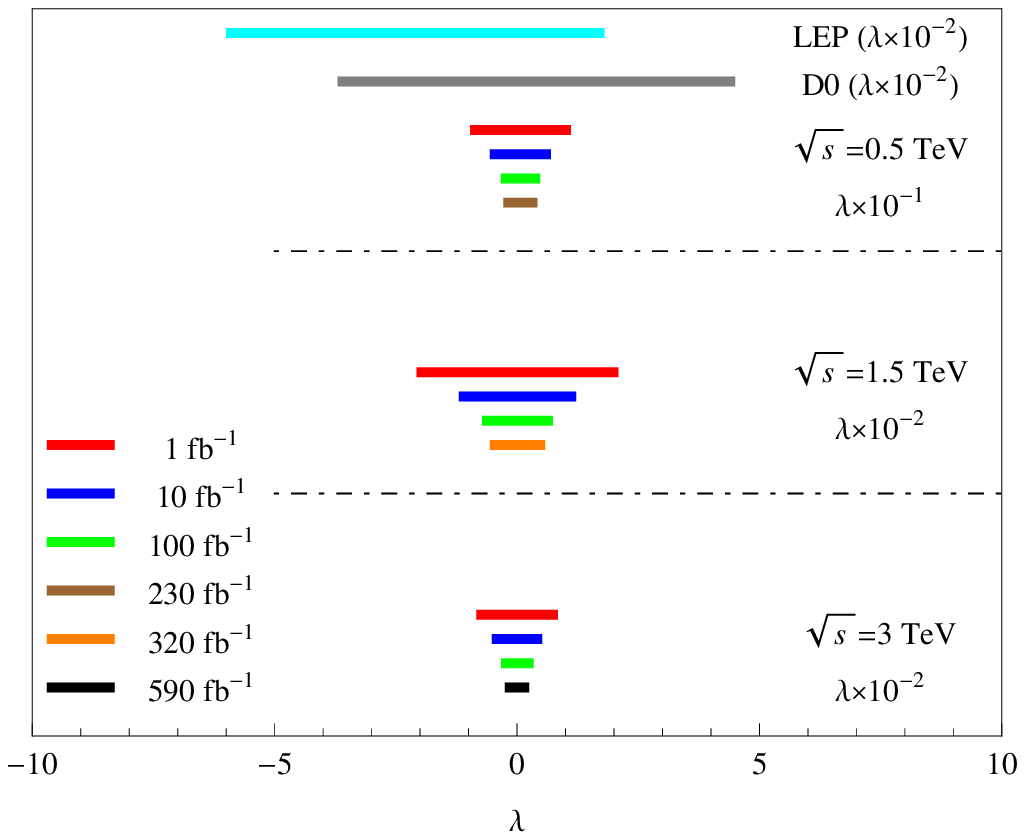}
\caption{95\% C.L. sensitivity limits of the $\lambda$ coupling
for various values of integrated CLIC luminosities and center-of-mass energies. $e^{-}e^{+}\rightarrow W^{-}W^{+} \to q\bar{q'}lv$ ($q;q'=u,d,s;l=e,\mu$) and processes with unpolarized beams have been considered.}
\label{figliml}
\end{figure}

\begin{figure}
\includegraphics{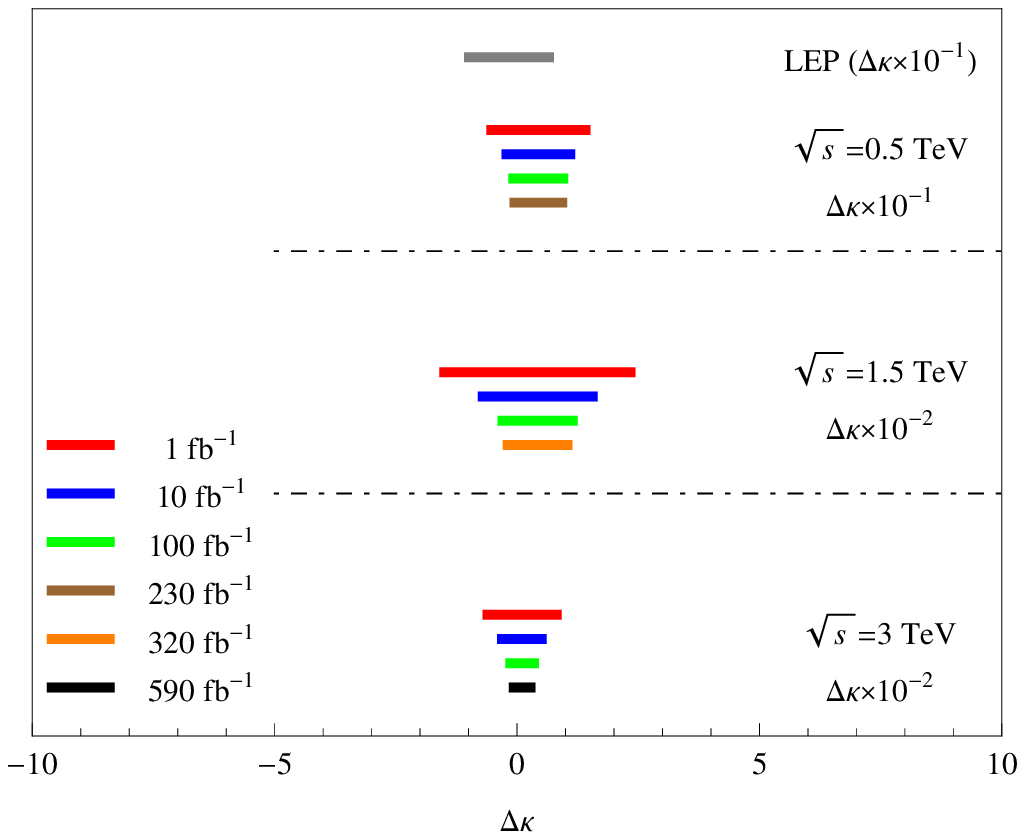}
\caption{95\% C.L. sensitivity limits of the $\Delta\kappa$ coupling
for various values of integrated luminosities and center-of-mass energies. $e^{-}e^{+}\rightarrow W^{-}W^{+} \to q\bar{q'}lv$ ($q;q'=u,d,s;l=e,\mu$) processes with $P(e^{-})=80\%$, $P(e^{+})=0\%$ polarizations have been considered.}
\label{figlimkpol1}
\end{figure}

\begin{figure}
\includegraphics{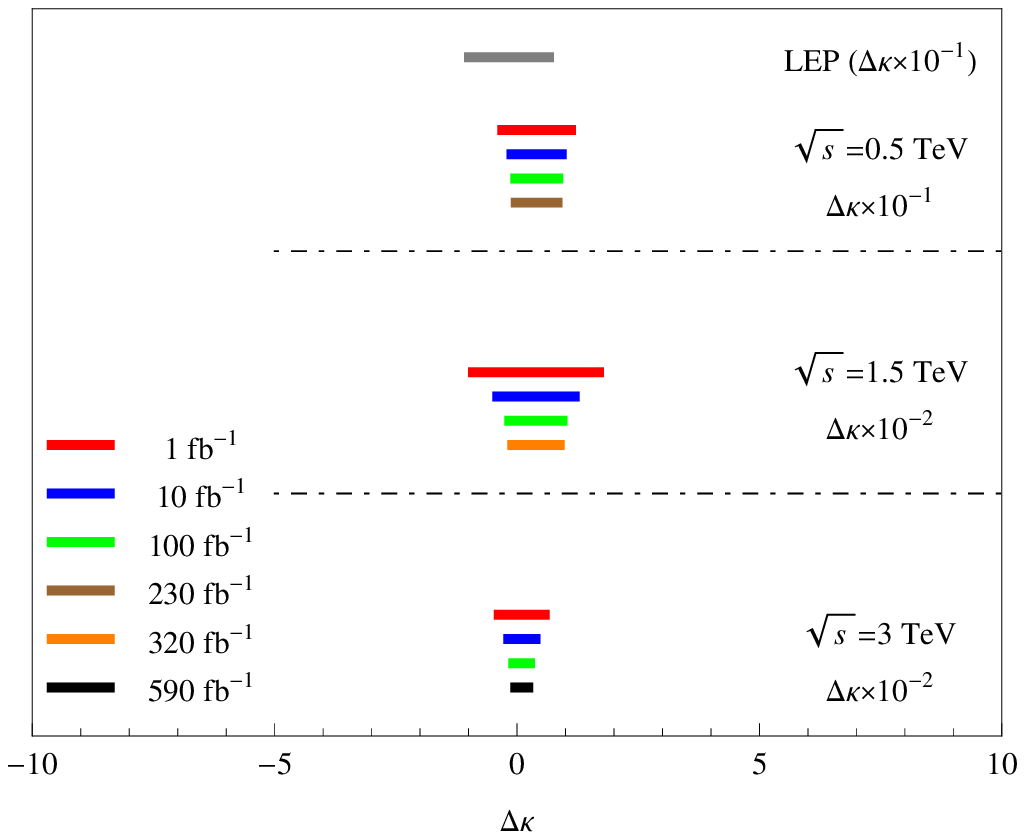}
\caption{95\% C.L. sensitivity limits of the $\Delta\kappa$ coupling
for various values of integrated luminosities and center-of-mass energies. $e^{-}e^{+}\rightarrow W^{-}W^{+} \to q\bar{q'}lv$ ($q;q'=u,d,s;l=e,\mu$) processes with $P(e^{-})=80\%$, $P(e^+)=$-$60\%$ polarizations have been considered.}
\label{figlimkpol2}
\end{figure}

\begin{figure}
\includegraphics{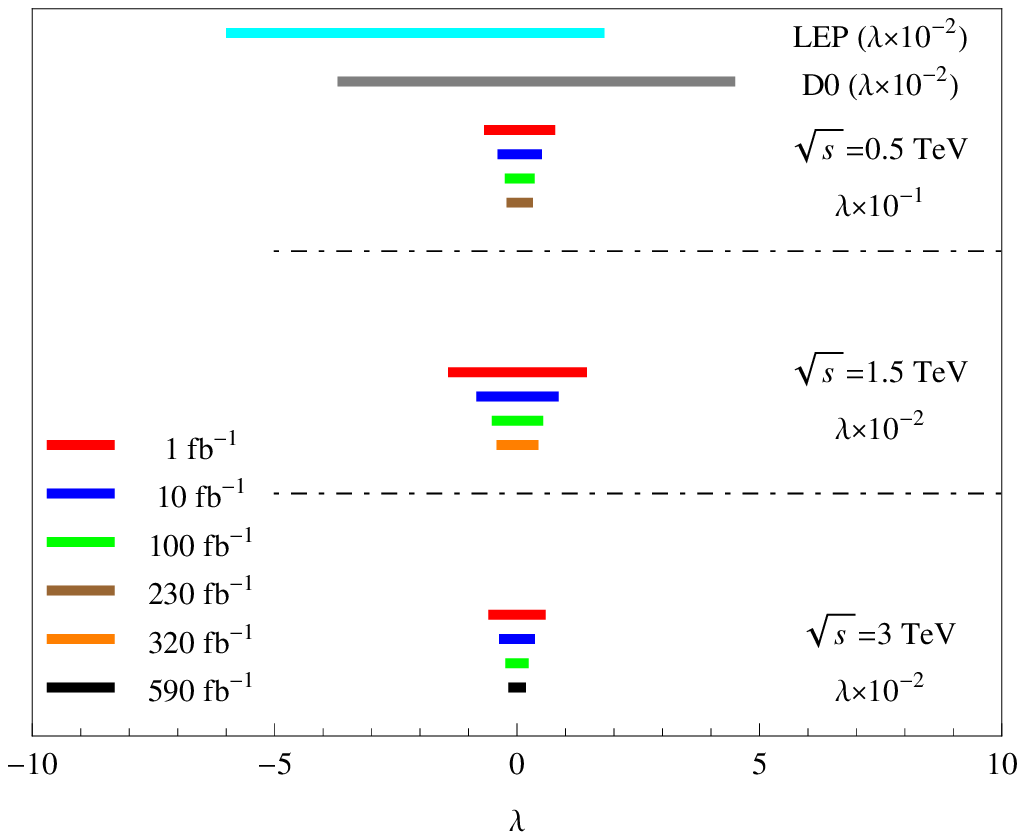}
\caption{95\% C.L. sensitivity limits of the  $\lambda$ coupling
for various values of integrated luminosities and center-of-mass energies. $e^{-}e^{+}\rightarrow W^{-}W^{+} \to q\bar{q'}lv$ ($q;q'=u,d,s;l=e,\mu$ processes with $P(e^{-})=80\%$, $P(e^{+})=0\%$ polarizations have been considered.}
\label{figlimlpol1}
\end{figure}

\begin{figure}
\includegraphics{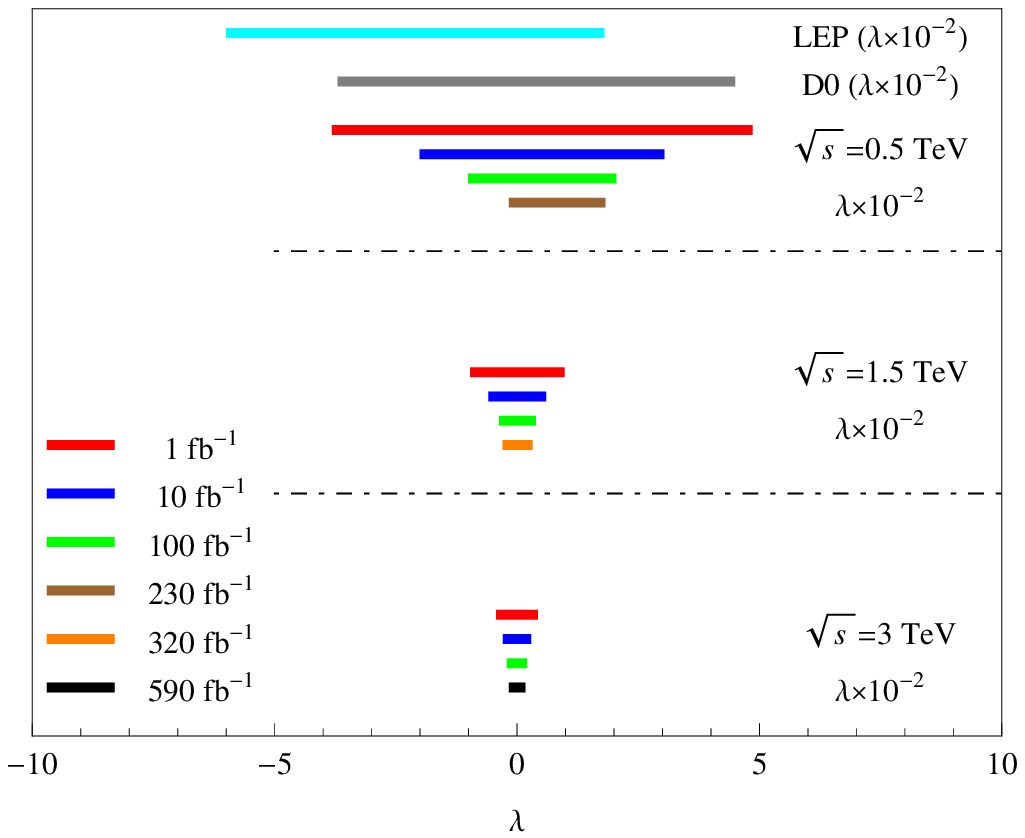}
\caption{95\% C.L. sensitivity limits of the $\lambda$ coupling
for various values of integrated luminosities and center-of-mass energies. $e^{-}e^{+}\rightarrow W^{-}W^{+}\to q\bar{q'}lv$ ($q;q'=u,d,s;l=e,\mu$ processes with $ P(e^{-})=80\%$, $P(e^+)=$-$60\%$ polarizations have been considered.}
\label{figlimlpol2}
\end{figure}

In Figs.\ref{figlimk} and \ref{figliml}, we have shown that limits on anomalous couplings for different luminosities and center-of-mass energies for the unpolarized case. We have seen from these figures that limits on the anomalous couplings are improved for increasing luminosities and center-of-mass energies. In addition to this result, our limits on $ \Delta\kappa$ anomalous coupling are weaker than $\lambda$ coupling because of the energy dependencies of these couplings as seen from Eq.\ref{efl}. It can be seen from Figs.\ref{figlimk} and \ref{figliml} that these limits are better with respect to the current experimental limits. We have also obtained the sensitivity limits of the anomalous couplings for different beam polarization cases. The limits obtained on $ \Delta\kappa$ anomalous coupling for different luminosities and center-of-mass energies with $P(e^-)=80 \%$, $P(e^+)=0\%$ and $P(e^-)=80 \%$, $P(e^+)=$-$60\%$ polarizations are presented in Figs.\ref{figlimkpol1} and \ref{figlimkpol2}, respectively. We can notice from these figures that polarization of the leptons can improve the limits on the anomalous couplings. In Figs.\ref{figlimlpol1} and \ref{figlimlpol2}, limits on the $\lambda$ anomalous coupling with beam polarization have been shown. Here, a similiar behavior for $ \Delta\kappa$ coupling has been observed.

\subsection{Anomalous couplings via the process $e^{-} \gamma^{*} \to W^{-}\nu_{e} $}

\begin{figure}
\includegraphics{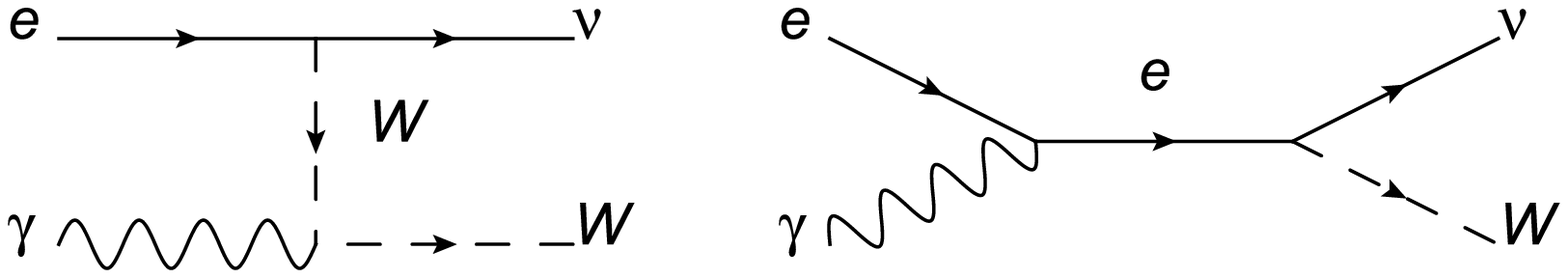}
\caption{Tree-level Feynman diagrams for the subprocess $e^{-} \gamma^{*} \rightarrow W^{-}\nu_{e} $ .
\label{figegdiagram}}
\end{figure}

In this subsection, we examine the process $e^{-} \gamma^{*} \rightarrow W^{-}\nu_{e}\rightarrow q q' \nu_{e}$ through the WWA approximation at the linear collider's energies. The $e\gamma$ interactions would be appropriate for examining heavy gauge boson
production processes \cite{che, che1}, since the incoming photon ensures us with a possibility
to detect directly the self-interactions of the gauge boson.
There are two Feynman diagrams for the process $e^{-} \gamma^{*} \rightarrow W^{-}\nu_{e}$ as seen from the
Fig.\ref{figegdiagram}. The first diagram allows the process suitable for examining the non-Abelian gauge structure of the theory
since it involves a triple gauge boson vertex.
An other important feature of this process is that it is sensitive both $\gamma WW$ and $ZWW$ couplings, therefore it is possible the distinguish the anomalous coupling of the photon and the neutral gauge boson $Z$.
The cross section of $e^{-} \gamma^{*} \rightarrow W^{-}\nu_{e}$ process approaches a constant value at
high energies in SM. Any signal that conflicts with the SM would
altered the good high energy behaviour and induce to a violation of unitarity at some energy.
At CLIC energies these deflection are expected to be small. Therefore, it will be difficult to detect this signal.
However, these effects can raise up using polarized electrons. This process has been investigated for the Compton backscattering photons in \cite{eg, eg1}.

\begin{figure}
\includegraphics{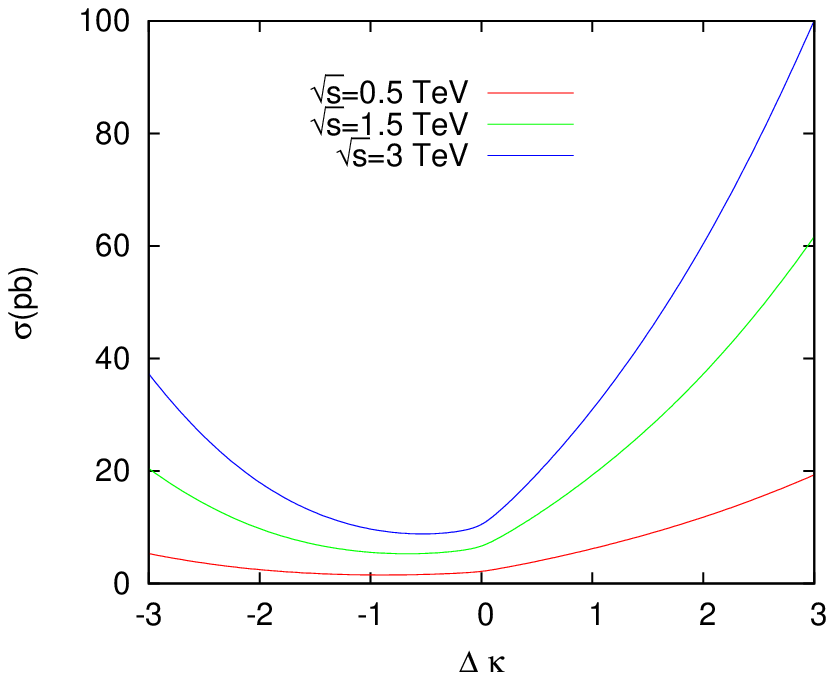}
\caption{The total cross section for $e^{-} \gamma^{*} \rightarrow W^{-}\nu_{e} $ processes as a function of $\Delta\kappa$ coupling at unpolarized case and various values of center-of-mass energy.
\label{figcs2k}}
\end{figure}

\begin{figure}
\includegraphics{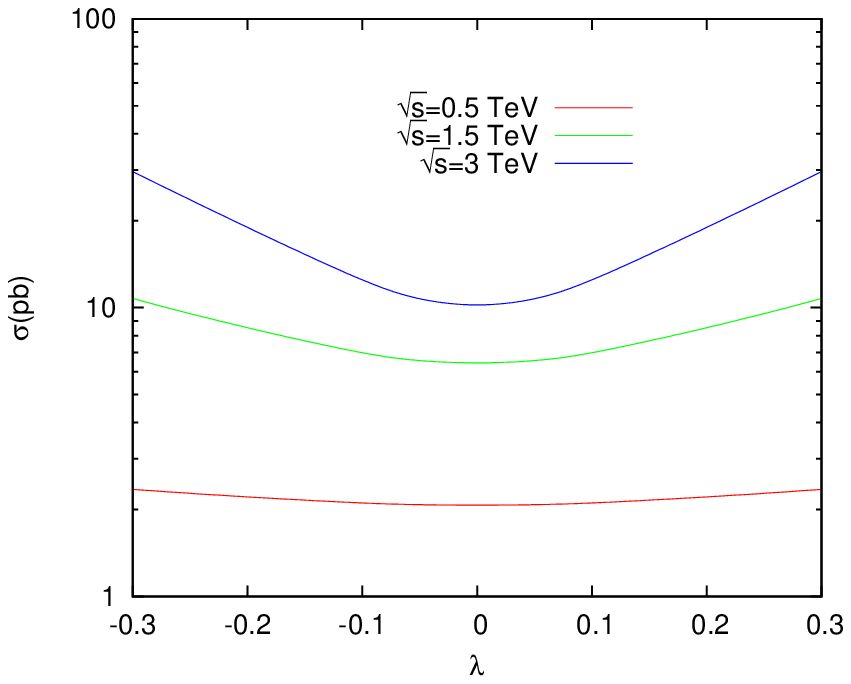}
\caption{The total cross section for $e^{-} \gamma^{*} \rightarrow W^{-}\nu_{e}$ processes as a function of $\lambda$ coupling at unpolarized case and various values of center-of-mass energy.
\label{figcs2l}}
\end{figure}

We show  the total cross section as a function of $\Delta\kappa$ and $\lambda$ anomalous couplings for three different center-of-mass energies at Figs.\ref{figcs2k} and \ref{figcs2l}. It can be seen from these figures that total cross section increases with center-of-mass energy. For $\Delta\kappa$ coupling, this increment is much less than $\lambda$ due to the energy dependence of the coupling.
The cross section is sensitive to the sign of the $\Delta\kappa$  as seen from the Fig.\ref{figcs2k}.
In the case of $500$ GeV center of mass energy, the cross-section is less sensitive to the sign of $\Delta\kappa$. Thus, we see from the in Fig.\ref{figcs2k} that deviation of the the cross section of $\Delta\kappa$ from linearity increases at  higher energies. In Fig.\ref{figcs2l} the cross section almost symmetric under the sign of the $\lambda.$ Therefore, main contribution comes from the $\lambda^2$ term of the cross section due to energy dependence of the $\lambda$.

\begin{figure}
\includegraphics{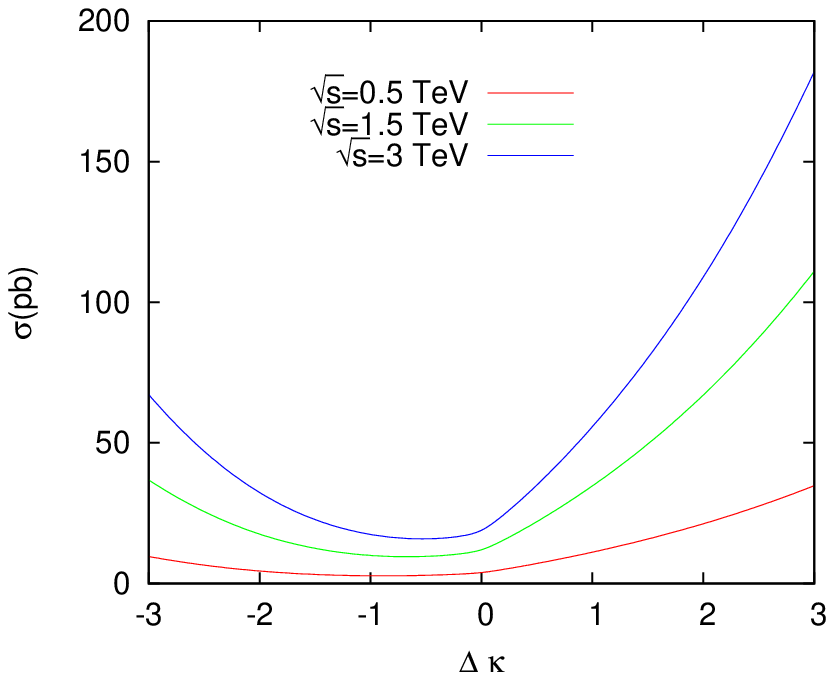}
\caption{The total cross section for $e^{-} \gamma^{*} \rightarrow W^{-}\nu_{e}$ processes as a function of $\Delta\kappa$ at the $\sqrt{s}=0.5, 1$ and $3$ TeV with $P(e^-)=$-$80\%$ polarization.
\label{figcs2kpol1}}
\end{figure}

\begin{figure}
\includegraphics{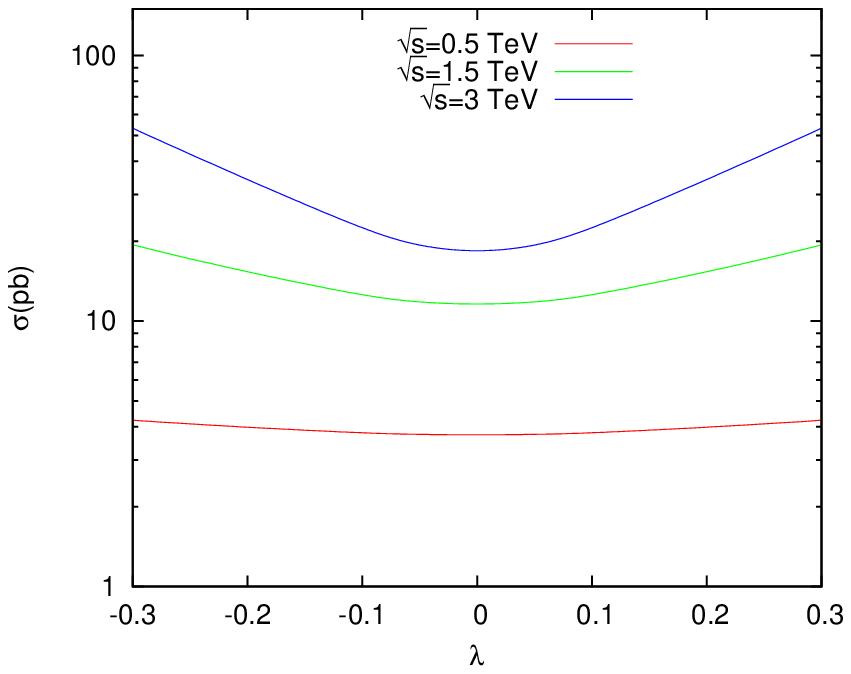}
\caption{The total cross section for $e^{-} \gamma^{*} \rightarrow W^{-}\nu_{e}$ processes as a function of $\lambda$ at the $\sqrt{s}=0.5, 1$ and $3$ TeV with $P(e^-)=$-$80\%$ polarization.
\label{figcs2lpol1}}
\end{figure}

\begin{figure}
\includegraphics{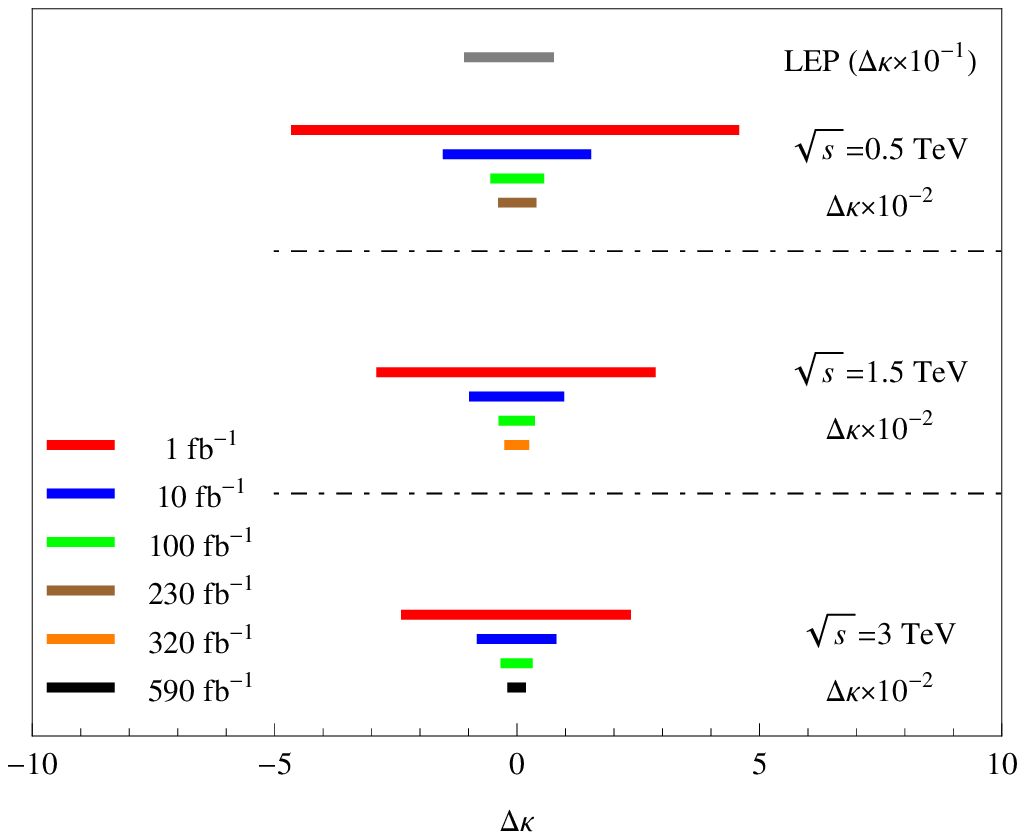}
\caption{95\% C.L. sensitivity limits of the $\Delta\kappa$ coupling
for various values of integrated luminosities and center-of-mass energies.$e^{-} \gamma^{*} \rightarrow W^{-}\nu_{e} \rightarrow q \bar{q'} \nu_{e}$ where $q;q'=u,d,s$, processes with unpolarized beams have been considered.}
\label{figlimk2}
\end{figure}

\begin{figure}
\includegraphics{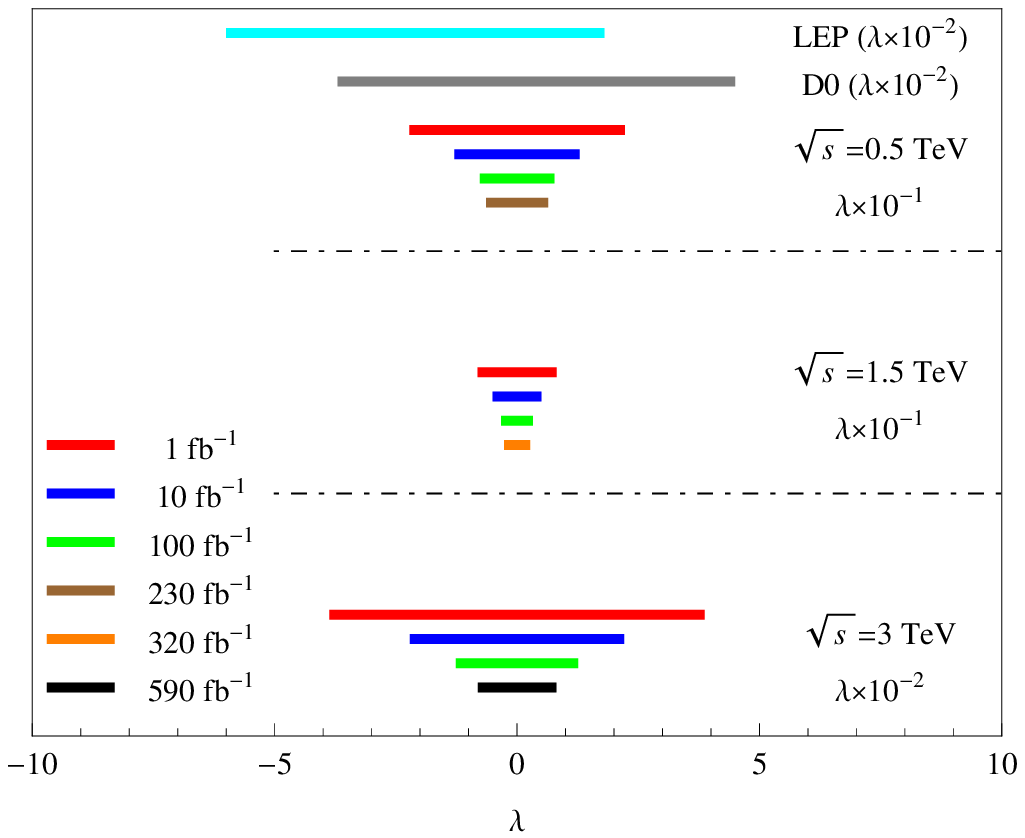}
\caption{95\% C.L. sensitivity limits of the  $\lambda$ coupling
for various values of integrated luminosities and center-of-mass energies.$e^{-} \gamma^{*} \rightarrow W^{-}\nu_{e} \rightarrow q \bar{q'} \nu_{e}$ $q;q'=u,d,s$, processes with unpolarized beams have been considered.}
\label{figliml2}
\end{figure}

\begin{figure}
\includegraphics{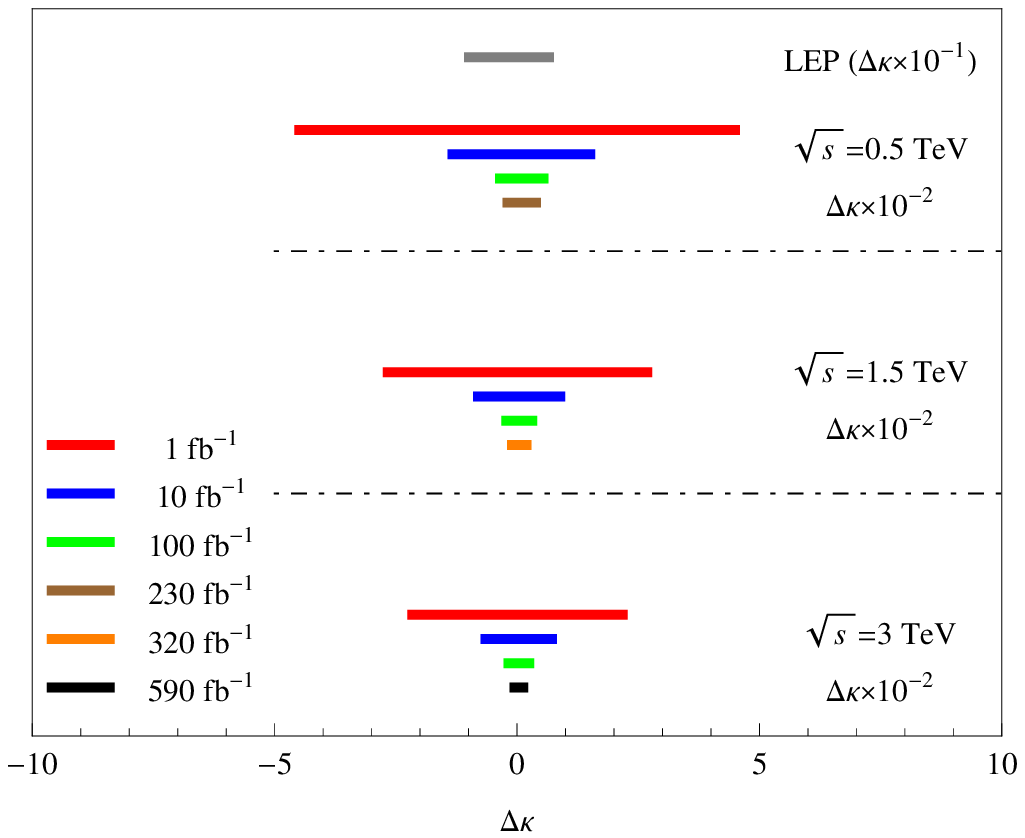}
\caption{95\% C.L. sensitivity limits of the  $\Delta\kappa$ coupling
for various values of integrated luminosities and center-of-mass energies.$e^{-} \gamma^{*} \rightarrow W^{-}\nu_{e} \rightarrow q \bar{q'} \nu_{e}$ where $q;q'=u,d,s$, processes with $P(e^-)=$-$80\%$ polarization have been considered.}
\label{figlimk2pol1}
\end{figure}

\begin{figure}
\includegraphics{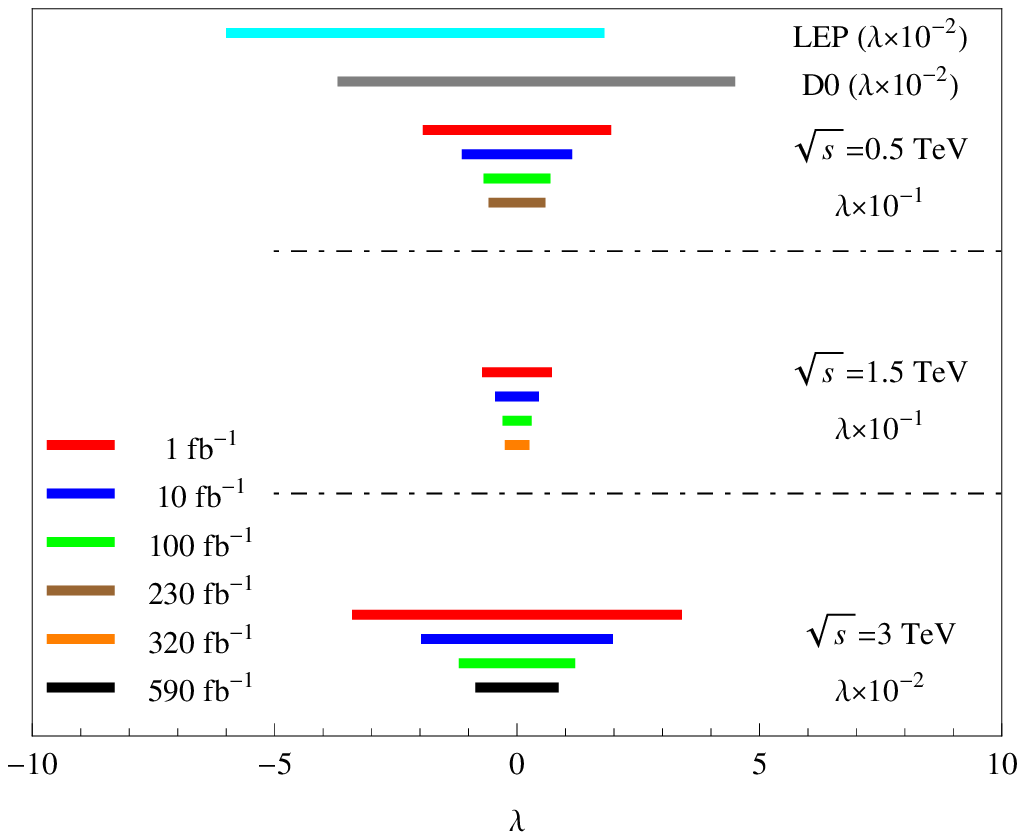}
\caption{95\% C.L. sensitivity limits of the   $\lambda$ coupling
for various values of integrated luminosities and center-of-mass energies.$e^{-} \gamma^{*} \rightarrow W^{-}\nu_{e} \rightarrow q \bar{q'} \nu_{e}$ where $q;q'=u,d,s$, processes with $P(e^-)=$-$80\%$ have been considered.}
\label{figliml2pol1}
\end{figure}

 We also examine these couplings for polarized case. The behaviour of the cross sections are not change as a function of $\Delta\kappa$ and $\lambda$. However, It is clear from  the Figs.\ref{figcs2kpol1} and \ref{figcs2lpol1} that the polarization ($P(e^-)=$-$80\%$) enhances the cross sections according to the unpolarized case.  The main reason of these results can be seen from Fig.\ref{figegdiagram}. There are two diagrams which contribute to the process and one of them includes $WW\gamma $ interaction vertex. This diagram gives the maximum contribution to the total cross section. For $P(e^-)=$-$80\%$ case, this contribution is dominant due to the structure of the $We\nu_{e}$ vertex. We have seen from Figs.\ref{figlimk2} and \ref{figliml2} that limits on the anomalous couplings are improved for increasing center-of-mass energies and these limits are better than the current experimental data. We also have shown from Figs.\ref{figlimk2pol1} and \ref{figliml2pol1}, there is no much effect of polarization on the limits.

\subsection{Anomalous couplings via the process $\gamma^{*} \gamma^{*} \rightarrow W^{-}W^{+}$}

\begin{figure}
\includegraphics{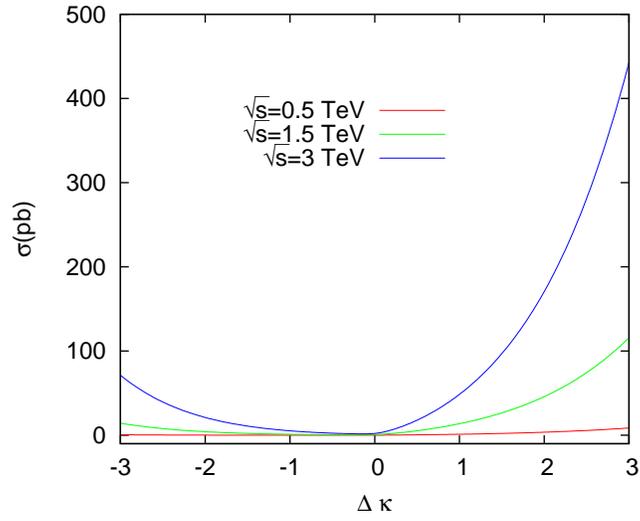}
\caption{The total cross section for  $ \gamma^{*} \gamma^{*} \rightarrow W^{-}W^{+}$ process as a function of $\Delta\kappa$ coupling at various values of center-of-mass energy.
\label{figcsk3}}
\end{figure}

\begin{figure}
\includegraphics{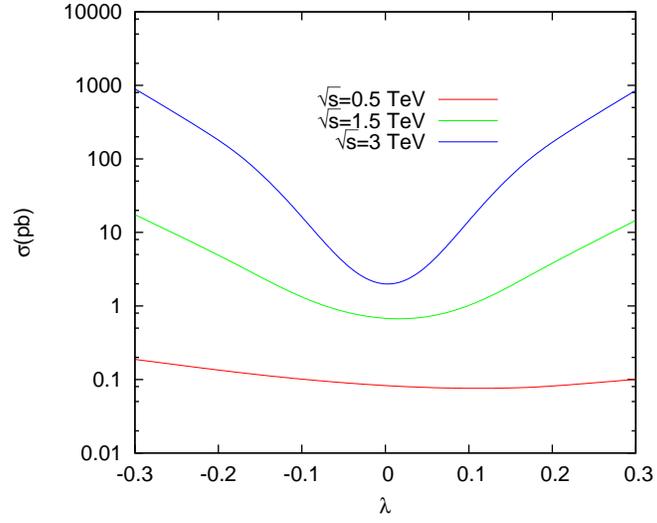}
\caption{The total cross section for $ \gamma^{*} \gamma^{*} \rightarrow W^{-}W^{+}$ process as a function of $\lambda$ coupling at various values center-of-mass energy.
\label{figcsl3}}
\end{figure}

\begin{figure}
\includegraphics{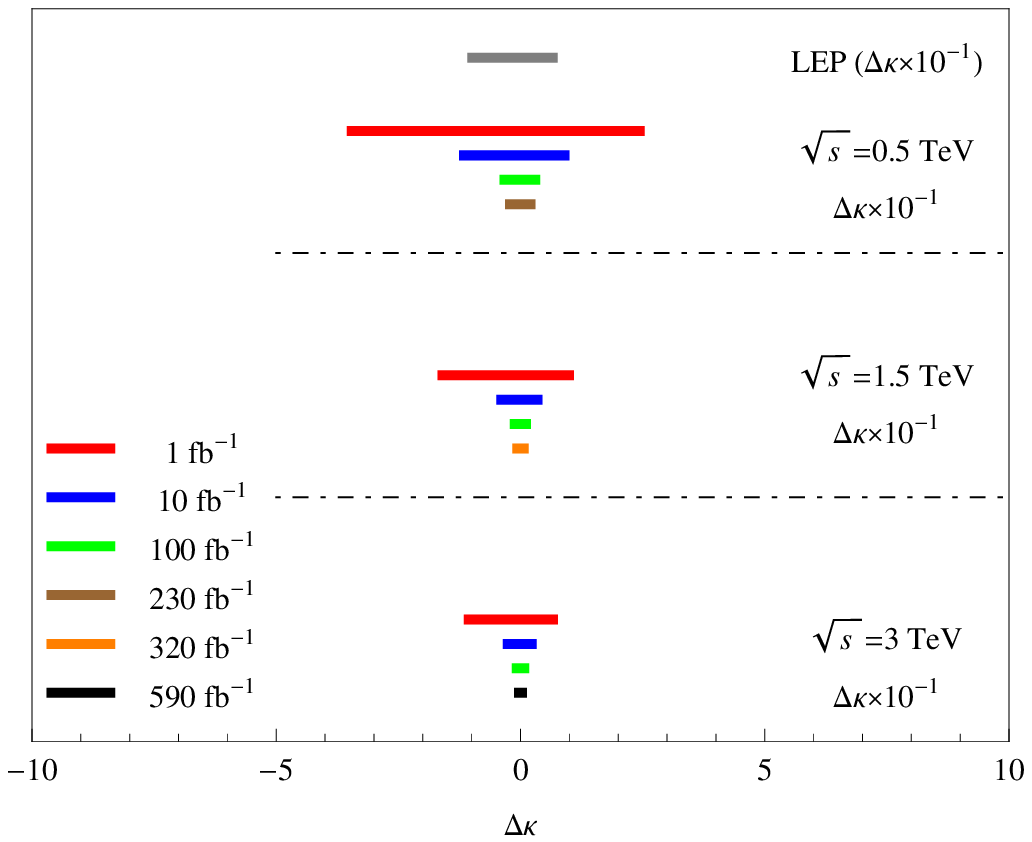}
\caption{95\% C.L. sensitivity limits of the  $\Delta\kappa$ coupling
for various values of integrated luminosities and center-of-mass energies. $\gamma^{*} \gamma^{*} \rightarrow W^{-}W^{+} \to q\bar{q'}lv$ ($q;q'=u,d,s;l=e,\mu$) processes with unpolarized beams have been considered.}
\label{figlimk3}
\end{figure}

\begin{figure}
\includegraphics{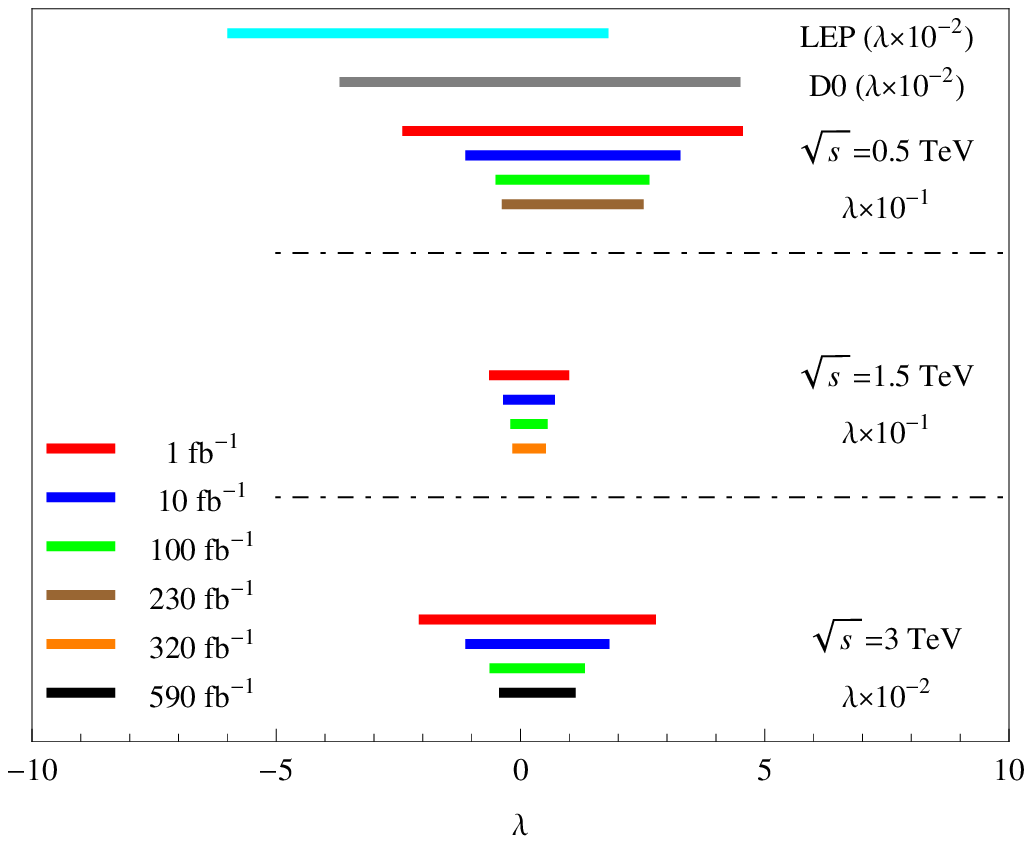}
\caption{95\% C.L. sensitivity limits of the  $\lambda$ coupling
for various values of integrated luminosities and center-of-mass energies. $\gamma^{*} \gamma^{*} \rightarrow W^{-}W^{+} \to q\bar{q'}lv$ ($q;q'=u,d,s;l=e,\mu$) processes with unpolarized beams have been considered.}
\label{figliml3}
\end{figure}

There is another contribution to $W^{-}W^{+}$ production via $\gamma^{*} \gamma^{*} \rightarrow W^{-}W^{+}\rightarrow (q q' l \nu_{l})$ process with $WW\gamma$ couplings.
The $e^+e^- \to W^+ W^-$ and $e^{-} \gamma^{*} \rightarrow W^{-}\nu_{e}$ processes includes only $3$-
boson interactions. Also specific $WW\gamma\gamma$ vertex is predicted in SM. That vertex contributes
to $\gamma^{*} \gamma^{*} \rightarrow W^{-}W^{+}$ making it a particularly important tool
in searching W electromagnetic interactions.
 For this process, we have drawn total cross section as a function of anomalous parameters in Figs.\ref{figcsk3} and \ref{figcsl3}. Changes according to the anomalous couplings of the cross sections are similiar to the previous processes. This process is found to be quite sensitive to anomalous
couplings such as $e^+e^- \to W^+ W^-$ and $e^{-} \gamma^{*} \rightarrow W^{-}\nu_{e}$.
We have seen from Figs.\ref{figlimk3} and \ref{figliml3} that limits on the anomalous couplings are improved for increasing center-of-mass energies and these limits are better than the current experimental data.

Moreover, we have found the sensitivity limits on model parameters both for the Circular Electron Positron Collider (CEPC) and International Linear Collider (ILC). The small Higgs mass ($\thicksim125 GeV$) allows a Circular Electron Positron Collider (CEPC) as a Higgs Factory. CEPC could be achieved
the CEPC $250$ GeV center of mass energy with integrated luminosity $5000$ fb$^{-1}$. ILC could be achieved $500$ GeV center of mass energy with integrated luminosity 500 fb$^{-1}$. The precision of the measurements that can be made at the these colliders make possible us to estimate at what energy new physics may detect. Our results have been shown for these collider parameters in Tab.\ref{tab3} and Tab.\ref{tab4}.

\section{Conclusion}

$\gamma^{*} \gamma^{*}$ and $\gamma^{*} e^{-}$ collisions can be accomplished easily at linear colliders with no additional equipments. Although the $e^{-}e^{+}\rightarrow W^{-}W^{+}$ process gives the best limits on anomalous $WW\gamma$ couplings, this process has disadvantage due to the including anomalous $WWZ$ couplings at the tree level. Especially, $\gamma^{*} \gamma^{*}$ and $\gamma^{*} e^{-}$ collisions may provide a good opportunity to investigate purely anomalous  $WW\gamma$ couplings since it has cleaner background. In this work, we also have shown that beam polarization can improve the limits on the anomalous couplings.
\pagebreak

\pagebreak

\begin{table}
\caption{Experimental limits at $95\%$ C. L. on $\Delta\kappa$ and $\lambda$.
\label{tab1}}
\begin{ruledtabular}
\begin{tabular}{cccccc}
&ATLAS&CMS&D0&CDF&LEP \\
\hline
$\Delta\kappa$&${[-0.135;0.190]}$&${[-0.210;0.220]}$&${[-0.158;0.255]}$&${[-0.460;0.390]}$&${[-0.099;0.066]}$ \\
$\lambda$&${[-0.065;0.061]}$&${[-0.048;0.037]}$&${[-0.036;0.044]}$&${[-0.180;0.170]}$&${[-0.059;0.017]}$
\end{tabular}
\end{ruledtabular}
\end{table}

\clearpage

\begin{figure}
\includegraphics{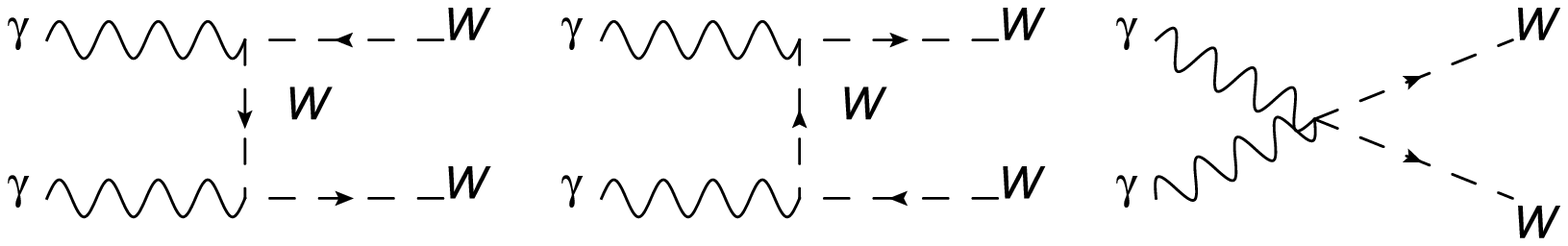}
\caption{Tree-level Feynman diagrams for the subprocess $\gamma^{*} \gamma^{*} \to W^+ W^- $ .
\label{figggdiagram}}
\end{figure}

\clearpage

\clearpage

\clearpage

\clearpage

\clearpage

\clearpage

\clearpage

\clearpage

\clearpage

\clearpage

\begin{table}
\caption{Experimental limits at $95\%$ C. L. on $\Delta\kappa$ and $\lambda$ at $\sqrt{s}=250$ GeV center of mass energy and $L=5000$ fb$^{-1}$.
\label{tab3}}
\begin{ruledtabular}
\begin{tabular}{cccc}
&$\lambda$ & $\Delta \kappa $ \\
\hline
$e^{-}e^{+}\rightarrow W^{-}W^{+}$ (unpolarized) &$[-0.00737 , 0.00755]   $&$[-0.00367 , 0.00347] $ \\
$e^{-}e^{+}\rightarrow W^{-}W^{+}$ ($P(e^{-})=+80\%$,$P(e^{+})=-60\%$) & $ [-0.00168 , 0.00173]   $ & $[-0.00102 , 0.00103] $ \\
$e^{-}e^{+}\rightarrow W^{-}W^{+}$ ($P(e^{-})=+80\%$,$P(e^{+})=0\%$)& $ [-0.00351 , 0.00367]   $ &  $ [-0.00206 , 0.00208]$ \\
\hline
\hline
$e^{-} \gamma \rightarrow W^{-} \nu_{e}$ (unpolarized) & $   [-0.0590 , 0.0590] $  &  $ [-0.00151 , 0.00156]$ \\
$e^{-} \gamma \rightarrow W^{-} \nu_{e}$ ($P(e^{-})=+80\%$) &  $  [-0.0863 , 0.0859]  $& $  [-0.00342, 0.00344]$\\
\hline
\hline
$\gamma \gamma \rightarrow W^{-}W^{+}$ (unpolarized) & $  [-0.0291,0.0334]  $& $ [-0.0171,0.0167]$\\
\end{tabular}
\end{ruledtabular}
\end{table}

\begin{table}
\caption{Experimental limits at $95\%$ C. L. on $\Delta\kappa$ and $\lambda$ at $\sqrt{s}=500$ GeV center of mass energy and $L=500$ fb$^{-1}$.
\label{tab4}}
\begin{ruledtabular}
\begin{tabular}{cccc}
&$\lambda$ & $\Delta \kappa $ \\
\hline
$e^{-}e^{+}\rightarrow W^{-}W^{+}$ (unpolarized) &$[-0.0139, 0.0280] $&$[-0.0057, 0.1305]$\\
$e^{-}e^{+}\rightarrow W^{-}W^{+}$ ($P(e^{-})=+80\%$,$P(e^{+})=-60\%$) & $[-0.0051, 0.0155]$ & $[-0.0018, 0.0828]$ \\
$e^{-}e^{+}\rightarrow W^{-}W^{+}$ ($P(e^{-})=+80\%$,$P(e^{+})=0\%$)& $[-0.0089, 0.0199] $ &  $[-0.0037, 0.0914]$ \\
\hline
\hline
$e^{-} \gamma \rightarrow W^{-} \nu_{e}$ (unpolarized) & $[-0.0443, 0.0444]$  &  $[-0.0019, 0.0020]$ \\
$e^{-} \gamma \rightarrow W^{-} \nu_{e}$ ($P(e^{-})=+80\%$) &  $[-0.0679, 0.0679]$& $ [-0.0059, 0.0061]$\\
\hline
\hline
$\gamma \gamma \rightarrow W^{-}W^{+}$ (unpolarized) & $[-0.0198, 0.2331]  $& $[-0.0144, 0.0138]$\\
\end{tabular}
\end{ruledtabular}
\end{table}

\end{document}